# Investigation of Shape Phase Transition in the $U(5) \leftrightarrow SO(6)$ Transitional Region by Catastrophe Theory and Critical Exponents of some Quantities

M. A. Jafarizadeh[a,b] [*], N. Fouladi[c], H. Fathi[c], M.Ghadami[c], H.sabri[c][†]

[a] Department of Theoretical Physics and Astrophysics, University of Tabriz, Tabriz 51664,Iran.

[b] Research Institute for Fundamental Sciences, Tabriz 51664, Iran.

[c] Department of Nuclear Physics, University of Tabriz, Tabriz 51664, Iran.

---

[*] E-mail: jafarizadeh@tabrizu.ac.ir

[†] E-mail: h-sabri@tabrizu.ac.ir

# Abstract

In the present paper, by employing the formation of the Catastrophe Theory, the phase transition points for $U(5) \leftrightarrow SO(6)$ transitional Hamiltonian, which is defined according to the affine $SU(1,1)$ algebra are investigated. The energy surfaces of the even-even isotopes of $Ru$ and $Pd$ nuclei are considered and special isotopes which are the best candidates for the $E(5)$ critical point are identified. Also, the analogy between the critical exponents of ground-state quantum phase transition and Landau values for the critical exponents of thermodynamic phase transition are described.

**Keywords**: Catastrophe Theory- coherent states- Interacting Boson Model (IBM) energy surface- phase transition point Bifurcation set- Critical Exponents.

## 1. Introduction

Atomic nuclei are known to exhibit changes of their energy levels and electromagnetic transition rates among them when the numbers of protons (or neutrons) are modified, resulting in quantum phase transitions from one kind of collective behavior to another. Quantum phase transition (QPT) occurs in the critical value of the control parameters and the zero temperature. The concept of QPT, sometimes called the ground state phase transition or phase transition at zero temperature, refers to the sudden change in the equilibrium states of the system. The state of a system is a function of state variables (order parameters) and control parameters. QPT in atomic nuclei occurs through the varying N or Z that causes the sudden change in the nuclear structure [1].This transitions were studied widely in the early 80's in the framework of the IBM. The general procedure was laid out by Gilmore in the 70's using atomic coherent states in the combination with Catastrophe Theory [2] which was applied to nuclear physics by Dieperink, Scholten and Iachello[3], Feng, Gilmore, Zhang and Deans[4-6] and L´opez-Moreno and Casta˜nos [7].In addition to theoretical methods, there are several experimental methods to study the phase transition in nuclei. For example, one can study $R_{4/2}$, two-neutron separation energies, E2 transition strengths, E0 transitions, isotope shifts, and two-nucleon transfer cross sections in order to empirically investigate QPT in nuclei [1].

In this paper, the $U(5) \leftrightarrow SO(6)$ transitional Hamiltonian which described by an affine $SU(1,1)$ algebraic technique is considered [8-9]. By using the Catastrophe theory [2] and considered Hamiltonian and intrinsic state of systems, we have generated the corresponding energy surface, in terms of the Hamiltonian parameters and the shape variables. To obtain the values of Hamiltonian's parameter which corresponds to the critical points, we analyze the energy surfaces and explore all the equilibrium configurations. The critical points of maximum degeneracy are identified where in our case, it corresponds to $\beta = 0$, and then, the bifurcation set are constructed. By applying the Catastrophe Theory in the considered Hamiltonian, some special isotopes which are the best candidates for $E(5)$ "critical symmetry" [10-12] are identified. Namely, different isotopes of $Ru$ and $Pd$ nuclei (which are knoen to located in this transitional region) are investigated where the $^{104}Ru$ and $^{102}Pd$ isotopes provide evidences for $E(5)$ critical symmetry. Also, the behavior of systems in the immediate neighborhood of the critical point by the critical exponents [13-14] are studied where the similarity between the critical exponents of ground-state quantum phase transition and the predictions of Landau for critical exponents of thermodynamic phase transition is verified.

The paper is structured as follows. In Sec2, we have reviewed the transitional Hamiltonian and main concepts. By employing the coherent states; we have evaluated the energy surface in the Sec3. Sec.4 represents the main concepts of elementary Catastrophe Theory while the results would be displayed in Sec.5. The analogy between critical exponents of ground-state quantum phase transition and Landau value for critical exponents of thermodynamic phase transition are then discussed in Sec.6. Finally, some conclusion and summary will be discussed in Sec.7.

## 2. The $U(5) \leftrightarrow SO(6)$ Transitional Hamiltonian

The Interacting Boson Model (IBM) based on general algebraic group theoretical techniques has been rather successful in describing the low-lying collective properties of nuclei a cross a wide range of nuclei. This model offers a simple Hamiltonian which describes the collective states by a system of interacting s- and d-bosons carrying angular momentum 0 and 2 respectively which define an overall $U(6)$ symmetry [15-16]. The IBM Hamiltonian provide exact solutions in three $U(5), SU(3)$ and $O(6)$ dynamical symmetry limits of theory which correspond to anharmonic vibrator, the axial rotor and $\gamma -$ unstable rotor as geometrical analogs, respectively. Another situation of complexity happens where the Hamiltonian considered in terms of invariant operator of these chains of symmetries which is considered as shape phase transition between these dynamical symmetry limits. The analytic description of the structure at the critical point of the phase transition where an obvious change happened in the structure is under investigations.

Recently, Iachello proposed the concept of "critical point symmetries" [10-11] in the Casten triangle [16] as presented in Figure1. This new class of symmetries has been provided for the systems localized at the critical points. In particular, the ''critical symmetry'' E(5) has been suggested to describe critical points in the phase transition from spherical to $\gamma$-unstable shapes while X(5) is designed to describe systems lying at the critical point in the transition from spherical to axially deformed systems [10-12]. Numerical methods must be employed to diagonalize the Hamiltonian in these cases. Pan et al [8-9] proposed a new solution which based on affine $SU(1,1)$ algebraic technique which prepares the properties of nuclei classified in the $U(5) \leftrightarrow SO(6)$ transitional region of IBM.

The $SU(1,1)$ Algebra has been described in detail in Ref [8-9]. Here, we briefly outline the basic ansatz and summarize the results. With using the $SU(1,1)$ generators, one can construct the following IBM Hamiltonian for the $U(5) \leftrightarrow SO(6)$ transitional case [8-9]

$$\hat{H} = g\, S_0^+ S_0^- + \alpha\, S_1^0 + \gamma\, \hat{C}_2(SO(5)) + \delta\, \hat{C}_2(SO(3)) \quad , \tag{2.1}$$

Where $g, \alpha, \gamma$ and $\delta$ are real parameter. The $C(SO(5))$ and $C(SO(3))$ represent the Casimir operators belonging to $SO(5)$ and $SO(3)$ groups, respectively. In the following, we introduce the $SU(1,1)$ algebra which the Hamiltonian is written in terms of its generators. This algebra is generated by $S^\mu$ where $\mu$ can take values 0 and $\pm$, which satisfies the following commutation relations,

$$[S^0, S^\pm] = \pm S^\pm \quad , \quad [S^+, S^-] = -2S^0 \tag{2.2}$$

In the IBM, d-boson pairing algebra generated by

$$S^+(d) = \frac{1}{2}(d^\dagger . d^\dagger) \quad , \quad S^-(d) = \frac{1}{2}(d.d) \tag{2.3a}$$

$$S^0(d) = \frac{1}{4}\sum_\mu (d^\dagger_\mu . d_\mu + d_\mu . d^\dagger_\mu) \quad , \tag{2.3b}$$

Similarly, the s-boson pairing algebra generated by

$$S^+(s) = \frac{1}{2}(s^\dagger . s^\dagger) \quad , \quad S^-(s) = \frac{1}{2}(s.s) \tag{2.4a}$$

$$S^0(s) = \frac{1}{4}(s^\dagger s + s s^\dagger) \quad , \tag{2.4b}$$

Now, consider the infinite dimensional algebra generated by the following relations [8-9]

$$[S_m^+, S_n^-] = -2 S_{m+n+1}^0 \quad , \quad [S_m^0, S_n^\pm] = \pm S_{m+n}^\pm \tag{2.5}$$

Where $c_s$ and $c_d$ are the real parameters, and $n$ can be taken $0, \pm 1, \pm 2, \ldots$. We can easily show that the generators, satisfy the following commutation relations

$$S_n^\pm = c_s^{2n+1} S^\pm(s) + c_d^{2n+1} S^\pm(d) \quad , \quad S_n^0 = c_s^{2n} S^0(s) + c_d^{2n} S^0(d) \tag{2.6}$$

Now, we rewrite the Hamiltonian terms by using (2.3) up to (2.6)

$$S_n^\pm = c_s^{2n+1} S^\pm(s) + c_d^{2n+1} S^\pm(d) \quad \Rightarrow \quad S_0^\pm = c_s S^\pm(s) + c_d S^\pm(d) \tag{2.7}$$

For the first term of the Hamiltonian, we have

$$g S_0^+ S_0^- = \frac{g}{4}(c_s^2 s^\dagger s^\dagger s s + c_s c_d s^\dagger s^\dagger (d.d) + c_s c_d (d^\dagger . d^\dagger) s s + c_d^2 (d^\dagger . d^\dagger)(d.d)) \quad , \tag{2.8}$$

And similarly for the second term

$$S_n^0 = c_s^{2n} S^0(s) + c_d^{2n} S^0(d) \quad \Rightarrow \quad \alpha S_1^0 = \alpha(\frac{c_s^2}{4}(s^\dagger s + s s^\dagger) + \frac{c_d^2}{4}\sum_\mu d^\dagger_\mu d_\mu + d_\mu d^\dagger_\mu) \quad , \tag{2.9}$$

So we have

$$H = \frac{g}{4}(c_s^2 s^\dagger s^\dagger s s + c_s c_d s^\dagger s^\dagger (d.d) + c_s c_d (d^\dagger . d^\dagger) s s + c_d^2 (d^\dagger . d^\dagger)(d.d)) +$$

$$\alpha(\frac{c_s^2}{4}(s^\dagger s + s s^\dagger) + \frac{c_d^2}{4}\sum_\mu d^\dagger_\mu d_\mu + d_\mu d^\dagger_\mu) + \gamma C_2(SO(5)) + \delta C_2(SO(3)) \quad , \tag{2.10}$$

It can be easily seen that (2.10) is equivalent to the $SO(6)$ Hamiltonian when $c_s = c_d$, and to the $U(5)$ Hamiltonian when $c_s = 0$ & $c_d \neq 0$. Therefore, the $c_s \neq c_d \neq 0$ just corresponds to the $U(5) \leftrightarrow SO(6)$ transitional situation. $SO(5)$ and $SO(3)$ Casimir operators can be written in terms of the creation and annihilation boson operators in the following form [17]

$$C_2(SO(5)) = G^3 . G^3 + G^1 . G^1 \quad , \quad C_2(SO(3)) = G^1 . G^1 \tag{2.11}$$

Where $G_l^k$ are Tensor operators, namely;

$$G^3 = [d^\dagger \times d]_k^3 \quad , \quad G^1 = [d^\dagger \times d]_k^1$$

Therefore, we have

$$C_2(SO(3)) = [d^\dagger \times d]^1_k \cdot [d^\dagger \times d]^1_k \quad , \tag{2.12}$$

$$C_2(SO(5)) = [d^\dagger \times d]^3_k \cdot [d^\dagger \times d]^3_k + [d^\dagger \times d]^1_k \cdot [d^\dagger \times d]^1_k \quad , \tag{2.13}$$

The tensor product of two tensors is written as

$$C_2(SO(5)) = \sum_{q_1=-3}^{3} (-1)^{q_1} [d^\dagger \times d]^3_{q_1} \cdot [d^\dagger \times d]^3_{-q_1} + \sum_{q_2=-1}^{1} (-1)^{q_2} [d^\dagger \times d]^1_{q_2} \cdot [d^\dagger \times d]^1_{-q_2} \quad , \tag{2.14}$$

Where

$$\sum_{q_1=-3}^{3} (-1)^{q_1} [d^\dagger \times d]^3_{q_1} [d^\dagger \times d]^3_{-q_1} =$$

$$= \sum_{q_1=-3}^{3} (-1)^{q_1} \sum_{m_1,m_2} \langle 2m_1 2m_2 | 3\, q_1 \rangle d^\dagger_{m_1} \tilde{d}_{m_2} \times \sum_{m_3,m_4} \langle 2m_3 2m_4 | 3\, -q_1 \rangle d^\dagger_{m_3} \tilde{d}_{m_4} \tag{2.15}$$

And

$$\sum_{q_2=-1}^{1} (-1)^{q_2} [d^\dagger \times d]^1_{q_2} [d^\dagger \times d]^1_{-q_2} =$$

$$= \sum_{q_2=-1}^{1} (-1)^{q_2} \sum_{m_1,m_2} \langle 2m_1 2m_2 | 3\, q_2 \rangle d^\dagger_{m_1} \tilde{d}_{m_2} \times \sum_{m_3,m_4} \langle 2m_3 2m_4 | 3\, -q_2 \rangle d^\dagger_{m_3} \tilde{d}_{m_4} \quad , \tag{2.16}$$

where the coefficients in the sum are ordinary Clebsh-Gordan (CG) coefficients.

## 3. Energy surfaces

The coherent state formalism of the IBM connects the algebraic and geometric descriptions of the three dynamical symmetry limits and also allows the study of the transitions between them. Using this formalism, one can evaluate the ground state energy as a function of shape variables $\beta$ and $\gamma$,. It was shown that for $U(5) \leftrightarrow SO(6)$ and $U(5) \leftrightarrow SU(3)$ phase transition take place. The main idea of the formulation of the condensate (coherent) states is based on considering the boson quadrupole pure states by the boson condensation as follows. The condensate states are used as trial wave functions for the zero-temperature variational procedure [18-19]

$$|N, \alpha_m\rangle = (s^\dagger + \sum_m \alpha_m d^\dagger_m)^N |0\rangle \quad , \tag{3.1}$$

Where $|0\rangle$ is the boson vacuum state, $s^\dagger$ and $d^\dagger$ are the boson operators of the IBM, and parameter $\alpha_m$ can be related to the deformation collective parameters of the quadrupole

$$\alpha_0 = \beta \cos \gamma \quad , \quad \alpha_{\pm 1} = 0 \quad , \quad \alpha_{\pm 2} = \frac{\beta}{\sqrt{2}} \sin \gamma \tag{3.2}$$

The inner product of the condensate states is as follows

$$\langle N, \alpha'_m | N, \alpha_m \rangle = N!(1 + \sum_m (\alpha'_m)^* \alpha_m)^N \quad , \tag{3.3}$$

And the effect of the operators on condensate states would be as follows

$$d|N, \alpha_m\rangle = N\alpha_m |N-1, \alpha_m\rangle \quad , \tag{3.4a}$$

$$s|N, \alpha_m\rangle = N|N-1, \alpha_m\rangle \quad , \tag{3.4b}$$

$$d^\dagger_m |N, \alpha_m\rangle = \frac{1}{N+1} \frac{\partial}{\partial \alpha_m} |N+1, \alpha_m\rangle \quad , \tag{3.4c}$$

$$s_m^\dagger |N,\alpha_m\rangle = (1 - \frac{1}{N+1}\alpha_m \frac{\partial}{\partial \alpha_m})|N+1,\alpha_m\rangle \quad , \tag{3.4d}$$

The equation is used to calculate the energy surface is [18]

$$E = \frac{\langle N,\alpha_m |H|N,\alpha_m\rangle}{\langle N,\alpha_m |N,\alpha_m\rangle} \quad , \tag{3.5}$$

Therefore, energy surfaces for each part of the Hamiltonian in terms of the state variables and control parameters is constituted by

$$<gS_0^+ S_0^-> = \frac{g}{4}(\frac{N(N-1)}{(1+\beta^2)^2})(c_s^2 + 2c_s c_d \beta^2 + c_d^2 \beta^4) \quad , \tag{3.6a}$$

$$<\alpha S_0^1> = \frac{\alpha c_s^2}{4}(\frac{2N}{1+\beta^2}+1) + \frac{\alpha c_d^2}{4}(\frac{2N\beta^2}{1+\beta^2}+5) \quad , \tag{3.6b}$$

$$<\gamma \hat{C}_2(SO(5))> = 2\frac{\gamma N \beta^2}{1+\beta^2} \quad , \tag{3.6c}$$

$$<\delta \hat{C}_2(SO(3))> = \frac{3}{5}\frac{\delta N \beta^2}{1+\beta^2} \quad , \tag{3.6d}$$

Consequently, the energy surfaces is given by

$$E(\beta,\gamma) = \frac{g}{4}(\frac{N(N-1)}{(1+\beta^2)^2})(c_s^2 + 2c_s c_d \beta^2 + c_d^2 \beta^4) + \frac{\alpha c_s^2}{4}(\frac{2N}{1+\beta^2}+1) + \frac{\alpha c_d^2}{4}(\frac{2N\beta^2}{1+\beta^2}+5) +$$
$$+2\frac{\gamma N \beta^2}{1+\beta^2} + \frac{3}{5}\frac{\delta N \beta^2}{1+\beta^2} \quad , \tag{3.7}$$

## 4. Catastrophe Theory

Quantum phase transition occurs at the critical value of the control parameter and at the zero temperature. There are many ways to study the phase transition in the nuclei. Studying the phase transition by the Catastrophe Theory, is the most powerful methods in this field [2]. This theory attempts to study how the qualitative nature of the solutions of equations depends on the parameters that appear in the equations, and provides the appropriate method for modeling the systems that are associated with the sudden changes. In this way, the first step is to find the critical points of the energy surfaces and determining whether they are Morse or Non-Morse. In the Morse point, energy surfaces can be approximated by a quadratic function, while in the Non-Morse points doing so is not possible and the energy surfaces are written as the critical functions which are composed of germ and perturbation. Perturbation in the vicinity of the non-critical or Morse point is ineffective in qualitative features of the system, while around a Non-Morse point, perturbation will change the nature of the system [2].

### 4.1) Quantum phase transition for the $U(5) \leftrightarrow SO(6)$ Hamiltonian

First, we obtain the critical points of the energy surfaces of equation (3.7) to determine the equilibrium structures. Among the critical points, point with the most degeneracy selected and called the fundamental root. Therefore

$$\frac{\partial E}{\partial \beta} = \frac{\beta}{(1+\beta^2)^3}[gN(N-1)(c_s+c_d)(c_d-c_s)\beta^2 + 2(\frac{N}{2}\alpha c_d^2 + 2\gamma N + \frac{3}{5}\delta N - \frac{N}{2}\alpha c_s^2)(1+\beta^2)] \quad , \tag{4.1}$$

From the above equation is found that $\beta = 0$ is a critical point for each selection of control parameters. Now, the energy surfaces would extend around this point, according to the Catastrophe Theory

$$E(\beta) = \frac{g}{4}N(N-1)c_s^2 + \frac{N}{2}\alpha c_s^2 + \frac{1}{4}\alpha(c_s^2 + 5c_d^2) +$$

$$+ \frac{1}{2}[N(N-1)gc_s(c_d - c_s) + N(\alpha(c_d^2 - c_s^2) + \frac{6}{5}\delta + 4\gamma)]\beta^2 +$$

$$+[\frac{3}{4}N(N-1)gc_s^2 - N(N-1)gc_sc_d + \frac{1}{4}N(N-1)gc_d^2 + \frac{1}{2}N\alpha(c_s^2 - c_d^2) - \frac{3}{5}N\delta - 2N\gamma]\beta^4$$

$$+O(5) + ... \quad , \tag{4.2}$$

The linear term in $\beta$ is not appearing because $\beta = 0$ is a critical point $(\partial E/\partial \beta = 0)$. The symbol $O(5)$ indicates terms of the order $\beta^5$ or higher that be ignored because in the energy surface with the $k$ control parameters, terms of the expansion to order $x^{k+2}$ is sufficient for describe the most general behavior of the energy surface [2].

**4.2 bifurcation set**

In the next step of the procedure, we determine the bifurcation set. A bifurcation set is the locus of the points in the space of essential control parameters at which a transition occurs from one local minimum to another. The bifurcation set are obtained by the condition $\det H = 0$ [2]. $H$ is the matrix of second derivates of the energy surface evaluated at the critical point,

$$H_{ij} = \frac{\partial^2 E(x_i, x_j)}{\partial x_i \partial x_j}\bigg|_{x_1^c, x_2^c} \quad , \tag{4.3}$$

The index c denotes that they are critical points. For facility in calculation by using the following change of variables we can rewrite (4.2) in the form

$$E(\beta) = A + A'\beta^2 + A''\beta^4 + ..... \quad , \tag{4.4a}$$

where

$$A = \frac{g}{4}N(N-1)c_s^2 + \frac{N}{2}\alpha c_s^2 + \frac{1}{4}\alpha(c_s^2 + 5c_d^2) \quad , \tag{4.4b}$$

$$A' = \frac{1}{2}[N(N-1)gc_s(c_d - c_s) + N(\alpha(c_d^2 - c_s^2) + \frac{6}{5}\delta + 4\gamma)] \quad , \tag{4.4c}$$

$$A'' = \frac{3}{4}N(N-1)gc_s^2 - N(N-1)gc_sc_d + \frac{1}{4}N(N-1)gc_d^2 + \frac{1}{2}N\alpha(c_s^2 - c_d^2) - \frac{3}{5}N\delta - 2N\gamma] \quad , \tag{4.4d}$$

From bifurcation set condition and that energy surface is a function of single variable, we have

$$\frac{1}{2}[N(N-1)gc_s(c_d - c_s) + N(\alpha(c_d^2 - c_s^2) + \frac{6}{5}\delta + 4\gamma)] = 0 \quad , \tag{4.4e}$$

So for $c_s$, we have

$$c_s = \frac{g(N-1)c_d + \sqrt{g^2(N-1)^2 c_d^2 + 4[g(N-1) + \alpha][\alpha c_d^2 + 4\gamma + \frac{6}{5}\delta]}}{2(g(N-1) + \alpha)} \quad , \tag{4.5}$$

As mentioned in previous sections, for different values of $c_s$ and $c_d$, Hamiltonian (2.10) can describe $U(5), SO(6)$ symmetries or $U(5) \leftrightarrow SO(6)$ transition case. In this paper, we assume $c_d = 1$ and to analyze the different symmetry states, evolution of $c_s$ would be considered $c_s = 0$ and $c_s = 1$ corresponding to $U(5)$ and $SO(6)$ symmetry limits, respectively [8-9]. Therefore we expect, $c_s = 0.5$, is the foremost for the phase transition between two symmetry limits ($E(5)$ critical point symmetry).

# 5. Results

By using the least square method to experimental data taken from [20-23], we estimated the other parameters appearing in the Hamiltonian for different isotopes. The estimated values for these quantities of $Ru$ isotopes are presented in Table1.

| Nuclei | $c_d$ | $g$ | $\alpha$ | $\gamma$ | $\delta$ |
|---|---|---|---|---|---|
| $Ru$ | 1 | 1 | 4.4770 | -1.2451 | -0.0237 |

Table 1: The parameters of Hamiltonian have used to obtain $c_s$ values

With using these parameters in (4.5), the following values for $c_s$ are evaluated where represented in Table2.

| Isotope | $^{100}_{44}Ru$ | $^{102}_{44}Ru$ | $^{104}_{44}Ru$ | $^{106}_{44}Ru$ | $^{108}_{44}Ru$ | $^{110}_{44}Ru$ |
|---|---|---|---|---|---|---|
| N | 6 | 7 | 8 | 9 | 10 | 11 |
| $c_s$ | 0.38 | 0.46 | 0.52 | 0.56 | 0.60 | 0.63 |

Table 2. The variation of control parameter ($c_s$) by the changes of N values for $Ru$ isotopes

From these tables, one can conclude, the $^{104}_{44}Ru$ exhibits foremost transitional behavior, similar to the prediction of [24-29]. Consequently, this nucleus is the best candidate for the $E(5)$ critical point.

On the other hand, we have concerned similar procedure for $Pd$ isotopes as following, namely, The estimated values for the quantities of Hamiltonian correspond to different $Pd$ isotopes are presented in Table3 and determined $c_s$ values are listed in Table4.

| nuclei | $c_d$ | $g$ | $\alpha$ | $\gamma$ | $\delta$ |
|---|---|---|---|---|---|
| $Pd$ | 1 | 1 | 4.0218 | -1.0013 | 0.0147 |

Table 3: The parameters of Hamiltonian have used to obtain $c_s$ values.

| isotope | $^{102}_{46}Pd$ | $^{104}_{46}Pd$ | $^{106}_{46}Pd$ | $^{108}_{46}Pd$ | $^{110}_{46}Pd$ | $^{112}_{46}Pd$ |
|---|---|---|---|---|---|---|
| N | 5 | 6 | 7 | 8 | 9 | 10 |
| $c_s$ | 0.51 | 0.56 | 0.60 | 0.64 | 0.67 | 0.69 |

Table 4. The variation of control parameter ($c_s$) by the changes of N values for $Pd$ isotopes

Similar to previous case, one can conclude, the $^{102}_{46}Pd$ exhibits foremost transitional behavior, as the prediction of [24-29]. Therefore, this nucleus is the best candidate for the $E(5)$ critical point.

## 5.1) Analysis of phase transition by energy surface diagrams

As mentioned in the previous section, the classical limit of the Hamiltonian can be used for investigations of phase transitions. Fortunato *et al* in Ref. [30] predict, the evolution of the energy surface goes from a pure $\beta^2$ to a combination of $\beta^2$ and $\beta^4$ that has a deformed minimum. At the critical point of this second order phase transition the energy surface is a pure $\beta^4$ that is approximated with a square well which is analytically solvable. It means, if we use qualitative energy surface diagrams, in the $U(5)$ limit, the minimum of energy occurs at the $\beta = 0$ and in the $SO(6)$ limit, the minimum is at the $\beta = 1$. Also, at the critical point of this phase transition, the energy has a flat behavior where $U(5)$ and $SO(6)$ symmetry limits explore harmonic oscillator and Mexican hat potential behaviors, respectively as have presented in Figure (2). On the other hand, if we consider the evolution of energy surfaces (4.2) versus $\beta$ for different isotopes of $Ru$ and $Pd$, similar behavior suggested where the $^{100}_{44}Ru\,(c_s = 0.38)$ and $^{110}_{44}Ru\,(c_s = 0.63)$ isotopes which describe closer behavior to the $U(5)$ and $SO(6)$ dynamical symmetry limits, exhibit similar variations of energy surfaces as have presented in Figure(3). Similar behavior of $Pd$ isotopes exhibited in Figure (4).

## 6. Critical exponents

For fluid system, as we become close to the critical point, some of the quantities of the system are related to the temperature as $f(T) \approx (T - T_c)^\beta$ for some exponents of $\beta$. The similar behaviors may be seen not as a function of temperature, but as a function of some other quantities of the system $f(x) \approx (x)^\beta$. These exponents ($\beta$) are called the critical exponents and naturally defined as $\lim_{x \to 0} \frac{\ln f}{\ln(\pm x)}$ [1]. As have been explained in detail in Ref.[13-14,31-32], some basic critical exponents in thermodynamics have been employed to describe the evolution of considered systems near the critical points while we would use the corresponds of them in our analysis of nuclear systems as have explained in the following..

The behavior of $E(A, A', A''; \beta)$ in (4.4a), near the critical point is determined by the signs of the coefficients $A', A''$. The coefficients $A', A''$ which are functions of $c_s$, are written as functions of the dimensionless quantity, $\hat{c}_s = \frac{c_s - c_{s-critical}}{c_{s-critical}}; c_{s-critical} = \frac{1}{2}$. The expansion of coefficient $A'$ around the $\hat{c}_s$ is

$$A' = \left\{ -\frac{1}{8}N(N-1)g - \frac{1}{4}N\alpha + \frac{1}{4}N(N-1)gc_d + N\alpha c_d^2 + \frac{6}{5}N\delta + 4N\gamma \right\}$$
$$+ \left[ \frac{1}{4}N(N-1)g(c_d - 1) - \frac{1}{2}N\alpha \right]\hat{c}_s + \left[ -\frac{1}{8}N(N-1)g - \frac{1}{4}N\alpha \right]\hat{c}_s^2 \quad , \tag{6.1}$$

And the coefficient $A'$, which changes signs at $c_{s-critical} = \frac{1}{2}$, is now written in terms of $\hat{c}_s$, as

$$A' = \left\{ -\frac{1}{8}N(N-1)g - \frac{1}{4}N\alpha + \frac{1}{4}N(N-1)gc_d + N\alpha c_d^2 + \frac{6}{5}N\delta + 4N\gamma \right\}$$
$$+ \left[ \frac{1}{4}N(N-1)g(c_d - 1) - \frac{1}{2}N\alpha \right]\hat{c}_s \quad , \tag{6.2}$$

And $A'$ is represented as a series in odd powers of $\hat{c}_s$ so that it's sign would change at $\hat{c}_s = 0$.
Similarly, the expansion of coefficient $A''$ around the $\hat{c}_s$ is

$$A'' = \left\{ \frac{3}{16}N(N-1)g + \frac{1}{8}N\alpha - \frac{1}{2}N(N-1)gc_d + \frac{1}{4}N(N-1)gc_d^2 - \frac{1}{2}N\alpha c_d^2 - \frac{3}{5}N\delta - 2N\gamma \right\}$$
$$+ \left[ \frac{1}{2}N(N-1)g(\frac{3}{4} - c_d) + \frac{1}{4}N\alpha \right]\hat{c}_s + \left[ \frac{3}{16}N(N-1)g + \frac{1}{8}N\alpha \right]\hat{c}_s^2 \quad , \tag{6.3}$$

The stable systems have $A'' > 0$ on both sides of $c_{s-critical} = 1/2$; therefore the series for $A''$ is represented as one of even powers of $\hat{c}_s$ or

$$A'' = \left\{ \frac{3}{16} N(N-1)g + \frac{1}{8} N\alpha - \frac{1}{2} N(N-1)gc_d + \frac{1}{4} N(N-1)gc_d^2 - \frac{1}{2} N\alpha c_d^2 - \frac{3}{5} N\delta - 2N\gamma \right\}$$
$$+ \left[ \frac{3}{16} N(N-1)g + \frac{1}{8} N\alpha \right] \hat{c}_s^2 \quad , \tag{6.4}$$

The condition for stability is that $E(A, A', A''; \beta)$ must be a minimum as a function of $\beta$. From Eq. (4.4a), this condition may be expressed as

$$\frac{\partial E}{\partial \beta} = 0 = 2A'\beta + 4A''\beta^3, \tag{6.5}$$

Where terms above $\beta^4$ are presumed negligible near $c_{s-critical} = 1/2$. For $c_s < 1/2$, the only real root is $\beta = 0$; for $c_s > 1/2$, the root $\beta = 0$ correspond to a local maximum, and therefore not to equilibrium. The other two roots are found to be

$$\langle \beta \rangle = \pm \left( -\frac{a_1' \hat{c}_s}{2C_2} \right)^{\frac{1}{2}} \quad , \tag{6.6}$$

Where

$$a_1' = \frac{1}{4} N(N-1)g(c_d - 1) - \frac{1}{2} N\alpha \quad , \tag{6.7a}$$

and

$$C_2 = \frac{3}{16} N(N-1)g + \frac{1}{8} N\alpha - \frac{1}{2} N(N-1)gc_d + \frac{1}{4} N(N-1)gc_d^2 - \frac{1}{2} N\alpha c_d^2 - \frac{3}{5} N\delta - 2N\gamma , \tag{6.7b}$$

Where only the first terms in the expansions for the $A'$ and $A''$ were used. Consequently, the analysis predicts, the equilibrium order parameter near the critical point should depend on the $\hat{c}_s$ as

$$\langle \beta \rangle = Cte(-\hat{c}_s)^{\frac{1}{2}} \quad , \tag{6.8}$$

which means, critical exponent for order parameter is $1/2$ ), similar to the prediction of Jolie *et al* in Ref.[1]. On the other hand, a very sensitive probe of the phase transitional behavior is the ground-state energy with respect to the control parameters [33-34],

$$C(\lambda_i)\big|_{\{\lambda_j\}} = -\frac{\partial^2}{\partial \lambda_i^2} \varepsilon_0(\vec{\lambda}) \quad , \tag{6.9}$$

(where all $\lambda_j$'s with $j \neq i$ are kept constant).

In the above discussed thermodynamic analogy, $\varepsilon_0(\vec{\lambda})$ is replaced by the equilibrium value of the thermodynamic potential $F_0(P,T)$. In our descriptions, by use of Eq. (4.4a), the ground-state energies are $A$, $A - \frac{A'^2}{4A''}$ for $\beta = 0$ and $\beta \neq 0$ respectively. From Eq. (6.9) the specific heats are,

$$C^+(\beta_0 = 0) = -\left[\frac{g}{2}N(N-1) + N\alpha + \frac{\alpha}{2}\right] \qquad , \qquad \text{for} \qquad c_s < \frac{1}{2} \qquad (6.10a)$$

$$C^-(\beta_0 \neq 0) = -\left[\frac{g}{2}N(N-1) + N\alpha + \frac{\alpha}{2}\right] + \frac{a_1'^2}{2C_2} \qquad , \qquad \text{for} \qquad c_s > \frac{1}{2} \qquad (6.10b)$$

These results propose no dependence of C on $\hat{c}_s$ either above or below $c_{s-critical} = 1/2$, therefore the values for the specific heat exponents are both zero. Also, this result suggests a discontinuity in the heat capacity in the phase transition point which in the agreement by Landau's theory. Furthermore, this discontinuity confirms the second order of phase transition takes place between $U(5) \leftrightarrow SO(6)$ limits. Also, one can conclude, in the $U(5) \leftrightarrow SO(6)$ transitional region, phase transition would occur in the single point ($c_{s-critical} = 1/2$ or $E(5)$ critical point) where in the first order phase transition, a coexistent region is located between dynamical symmetry limits.

In order to identify other critical exponents, according to the Landau theory, by use of Eq.(4.4a), the potential energy surface becomes as,

$$E(\beta) = A - h\beta + A'\beta^2 + A''\beta^4 + \dots \qquad , \qquad (6.11)$$

Where "$h\beta$" represents the contribution of the intensive parameter $h$ for points off the $h=0$ coexistence curve. The equilibrium equation of state is $\left(\partial E/\partial \beta\right)\big|_{\hat{c}_s, h} = 0$ which cause to (for any small $h$)

$$2(a_1'\hat{c}_s)\beta + 4(C_2)\beta^3 = h \qquad , \qquad (6.12)$$

On the other hand, it reduces to its former representation for $h = 0$. The susceptibility may be found as have introduced in Ref.[35], namely,

$$\chi_{c_s}^{-1} = (\frac{\partial h}{\partial \beta})\big|_{\hat{c}_s} = 2a_1'\hat{c}_s + 12C_2\beta^2 \qquad , \qquad (6.13)$$

For $c_s < 1/2$, $\beta = 0$ and $\chi_{c_s}^{-1} = 2a_1'\hat{c}_s$, which gives the critical exponent equal to 1. For $c_s > 1/2$, with $h = 0$, Eq.(6.12) gives $\beta^2 = -(a_1'\hat{c}_s)/2C_2$, and therefore $\chi_T^{-1} = 4a_1'(-\hat{c}_s)$ or the critical exponent equal to 1. Along the critical isotherm, (in the phase transition point) $\hat{c}_s = 0$, and $h = 4C_2\beta^3$ giving the critical exponent as the critical exponent equal to 3.

From these figures and tables, one can conclude a second order shape phase transition in the $U(5) \leftrightarrow SO(6)$ transitional region where the results of critical exponents for heat capacity reveal these predictions. Also, the $^{104}_{44}Ru$ and $^{102}_{46}Pd$ isotopes are considered as the best candidates for $E(5)$ critical point which describe special values of the energy ratios ($R_{4_1^+/2_1^+}$ rotational excitations and $R_{0_2^+/2_1^+}$ vibrational excitation) while are closer to the prediction of Iachello [10-11].

## 7. Summary

In this contribution, we have used the Catastrophe formalism to study the shape phase transition of the energy surfaces of $U(5) \leftrightarrow SO(6)$ transitional Hamiltonian based on affine $su(1,1)$ Lie algebra. In particular, the results are accomplished for isotopes of $Ru$ and $Pd$, where the special isotopes of each nuclei which are the best candidate for the $E(5)$ critical point are determined. The result suggests, the second-order phase transition between spherical and deformed shapes of atomic nuclei is an isolated point following from the Landau theory of phase transitions. Finally we disclose that Landau values for critical exponents are perfectly applicable to the energy functional of $U(5) \leftrightarrow SO(6)$ transitional Hamiltonian.

# Figure caption

Figure1. The extended Casten triangle. Transition points and their associated critical symmetries are indicated.

Figure2. Evolution of the potential energy surfaces in the U(5), SO(6) and transitional region $(U(5) - SO(6))$ respectively (from left to right).

Figure3: Potential energy surfaces in terms of deformation parameter ($\beta$) for the Hamiltonian (3.7) for different isotopes of Ru (N=6-11).

Figure4: Potential energy surfaces in terms of deformation parameter ($\beta$) for the Hamiltonian (3.7) for different isotopes of Pd (N=5-10).

**Figure1.**

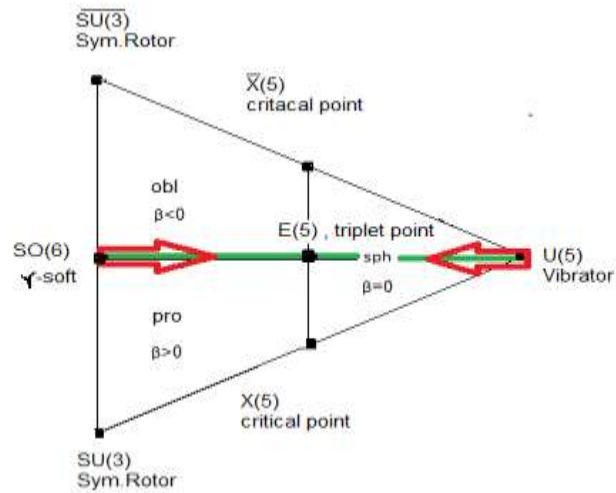

**Figure2.**

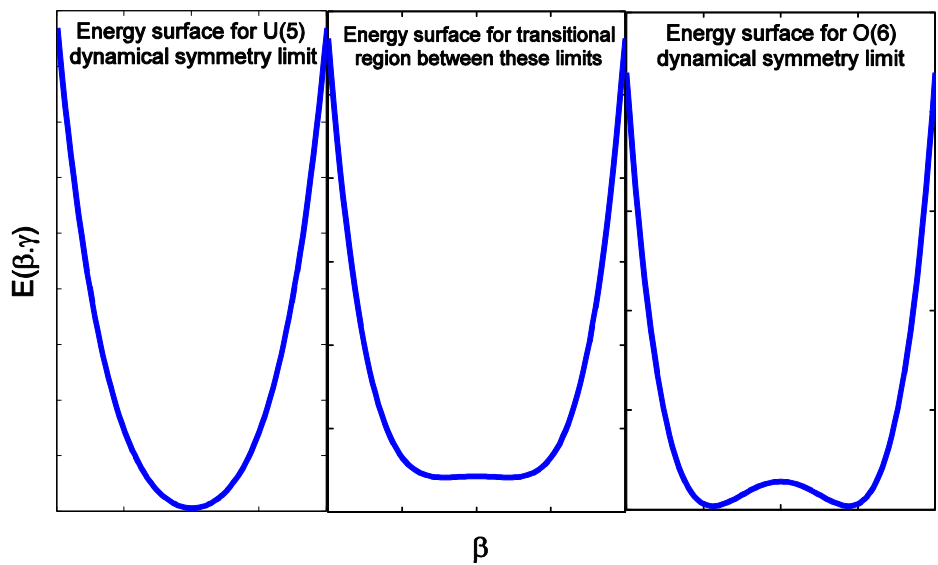

**Figure3.**

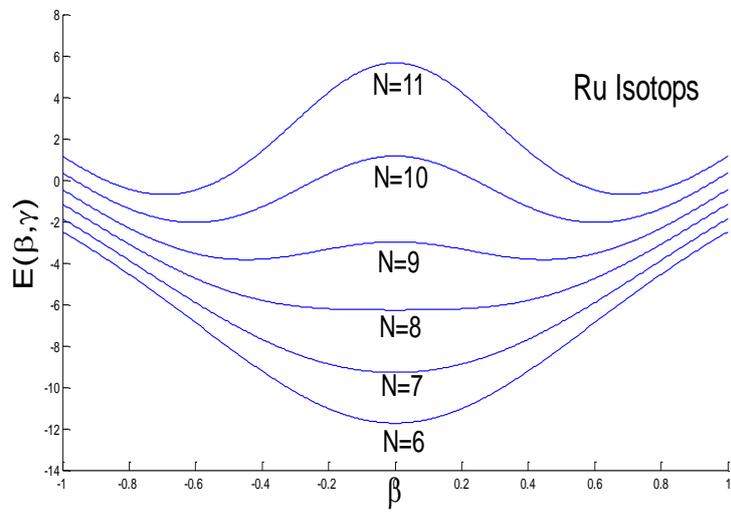

**Figure4.**

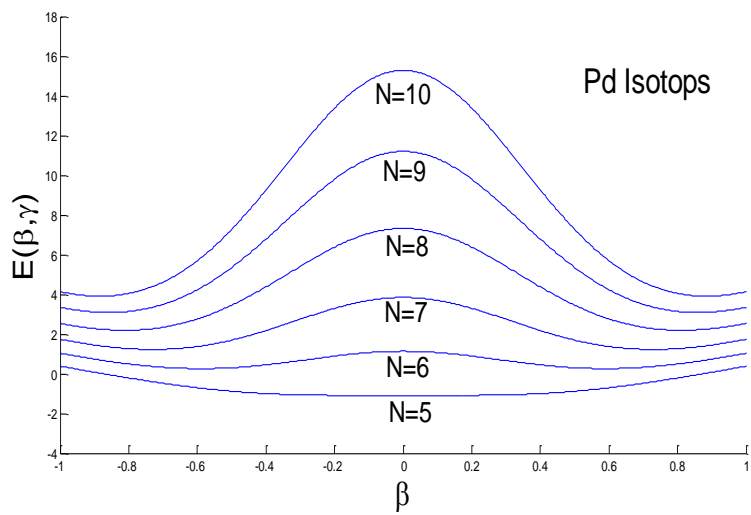

15